\documentclass[11pt]{article}
\usepackage[dvips]{graphicx}
\begin{document}

\begin{center}
\textbf{Rate equations for a Na+  channel gating master equation during the action potential within a neural membrane}
 \end{center}
\begin{center} S. R. Vaccaro  \end{center}
\begin{center} 
{\em Department of Physics, University of Adelaide, Adelaide, South Australia, 
5005, 
Australia} \\
  \end{center}
{\em svaccaro@physics.adelaide.edu.au} \\

\begin{quotation}
The action potential in a neural membrane is generated by Na+  and K+ channel ionic currents that may be calculated from a current equation and the 
 rate equations for activation variables $m$ and $n$, and the Na+ inactivation variable h. Assuming that a Na+ channel has three activation sensors, 
and  activation and inactivation are cooperative processes, a twelve state master equation that describes channel gating  may be reduced to 
kinetic equations for a five state system when the occupational probability of the first inactivated state is small, and the remaining inactivated states contribute to
 a  total inactivated state.  In the case of  independent activation sensors,    the inactivation rate is, in general,  dependent on the activation variable $m(t)$ as well as
the forward inactivation  transition rates. However,   when  $m(t)$   has a faster time constant than $h(t)$,   the inactivation rate   may be approximated by a
 voltage-dependent  function, and therefore,  the  solution of the master equation during an action potential may be approximated by the solution of 
Hodgkin-Huxley rate equations for   $m$  and   $h$.
\end{quotation}
\newpage

    {\bf INTRODUCTION} 

 By assuming that Na+ channel activation is  independent of inactivation, the Hodgkin-Huxley (HH)  rate equations for Na+ and K+ channels and the membrane current equation
 provide a good account of the action potential waveform, the threshold potential and subthreshold oscillations \cite{hh}, and the approach has been applied to a wide range of
 voltage-dependent ion channels in nerve, muscle and cardiac membranes. The HH rate equations for the Na+ channel are exact solutions of an eight state master equation for
 channel gating where the inactivation sensor and three activation sensors are independent \cite{hille,keener}. 
However, subsequent experimental studies have shown that the probability of Na+ inactivation increases with the degree of activation of the channel \cite{ab},
 the recovery from inactivation is more probable following deactivation \cite{kb}, and the kinetic equations for coupled Na+  activation and inactivation processes
represent ion channel states and their transitions, and provide a good description of the ionic and gating currents during a voltage clamp \cite{cgabc}.
From the solution of a nine state master equation that describes Na+ channel gating with two independent activation sensors and a coupled inactivation process,
 the open state probability during a depolarizing clamp may be expressed as $m^2 h$ where m(t) and h(t) satisfy rate equations, and therefore, the HH description of the
 Na+ current during a voltage clamp is consistent with a coupled Na+ channel gating model  \cite{vac1}.

 In this paper, it is shown that during an action potential, the solution of a six state master equation that describes coupling between a single 
Na+ channel activation sensor and a two-stage inactivation process may be approximated by interacting rate equations for activation and inactivation, when the first forward 
and backward inactivation  transitions are rate limiting.  When the Na+ channel has three activation sensors, by application of a reduction method, a twelve state master equation may be
 expressed as  kinetic equations for a five state system, and if the activation sensors are independent,  the inactivation rate is  dependent on the activation variable $m(t)$
 but may be approximated by a  voltage-dependent  function   when  $m(t)$   has a faster time constant than $h(t)$. Therefore,  the  solution of the Na+ channel master equation during an
 action potential may be approximated by  the solution of  HH rate equations for   $m$  and   $h$.

{\bf  Na+ CHANNEL MASTER EQUATION WITH A SINGLE ACTIVATION SENSOR} 

In this section, it is assumed that the activation of a single voltage sensor  regulating the  Na channel  conductance is coupled to a two-stage inactivation process (see Fig. 1),
and therefore, the kinetics may be described by a six-state master equation, that may be reduced to a four state system when the first forward and backward transitions are
 rate limiting \cite{vac1} (see Fig. 2)
\begin{eqnarray}
\frac{dC_1}{dt} & = &  -(\rho_1 + \alpha_m)C_1(t) + \beta_m O(t) + \sigma_1 B_1(t) \label{4m1} \\
\frac{dO}{dt} & = &  \alpha_m C_1(t) - (\beta_m + \rho_2) O(t) +  \sigma_2 B_2(t) \label{4m2} \\
\frac{dB_1}{dt} & = &  \rho_1 C_1(t) - (\alpha_{B} + \sigma_1) B_1(t) + \beta_{B} B_2(t)  \label{4m3} \\
\frac{dB_2}{dt} & = &  \rho_2 O(t) + \alpha_{B}B_1(t)  -  (\beta_{B} + \sigma_2) B_2(t)  \label{4m4}  
\end{eqnarray}
where $C_1(t)$, $O(t)$, $B_1(t)$, $B_2(t)$ are the occupational probabilities for the closed, open and inactivated states.

The Na current during deactivation of  an inactivated ion  channel is small  \cite{ab}, and therefore, the recovery rate $ \sigma_1 \gg \sigma_2 \approx 0$. It may be 
assumed that $\rho_1 \ll \alpha_{B}$  because inactivation is a slower process than activation, and during the action potential,  $ \sigma_1 \gg \beta_{B}$, and therefore,
 from Eq.  (\ref{4m3})
\begin{equation}	
B_1(t) \approx   \frac{ \rho_1 C_1(t) +  \beta_{B} B_2(t)}{\alpha_{B} + \sigma_1}.  
\end{equation}
That is, Eqs. (\ref{4m1}) and (\ref{4m4})  may be reduced to  (see Fig. 3)
\begin{eqnarray}
\frac{dC_1}{dt} & = &  -(\hat{\rho}_1 + \alpha_m)C_1(t) + \beta_m O(t) + \hat{\sigma}_1 B_1(t) \label{3m1} \\
\frac{dO}{dt} & = &  \alpha_m C_1(t) - (\beta_m + \rho_2) O(t) +  \sigma_2 B_2(t) \label{3m2} \\
\frac{dB_2}{dt} & = &  \rho_2 O(t) + \hat{\rho}_1 C_1(t)  - (\hat{\sigma}_1 + \sigma_2) B_2(t)  \label{3m3}  
\end{eqnarray}
where
\begin{eqnarray}
\hat{\rho}_1    & \approx &  \frac{\rho_1 \alpha_{B}}{\alpha_{B} + \sigma_1},   \label{r1}   \\
\hat{\sigma}_1  & \approx & \frac{\sigma_1 \beta_{B}  }{\alpha_{B} + \sigma_1} \label{s1},  
\end{eqnarray}
\noindent and $C_1(t) + O(t) + B_2(t) = 1$.  From Eq. (\ref{s1}), for moderate depolarizations $\hat{\sigma}_1   \approx   \beta_{B}$,
but for large hyperpolarizing potentials, $\hat{\sigma}_1 = \sigma_1 \beta_{B}/(\alpha_{B} + \beta_{B} + \sigma_1)$ and 
hence $\hat{\sigma}_1$ attains a plateau value of $\sigma_1$ when $\beta_{B} \gg  \sigma_1$ \cite{vac1}. 

Assuming that  $C_1(t) = (1 - m(t))h(t)$, 
$O(t) = m(t)h(t)$, $B_2(t) = 1 - h(t)$ in Eqs. (\ref{3m1}) to (\ref{3m3}), the variables  m(t) and h(t) satisfy
\begin{eqnarray}
\frac{dm}{dt}   & = &  \alpha_m  - m(t)[\alpha_m + \beta_m +  (\rho_2 - \hat{\rho}_1)(1 - m(t))] + \nonumber \\
 		&   & [\sigma_2 - m(t)(\hat{\sigma}_1 + \sigma_2)](1/h(t) - 1),  \label{mm} \\
\frac{dh}{dt}   & = &   \hat{\sigma}_1 + \sigma_2 - h(t)[ \hat{\sigma}_1 + \sigma_2 + \hat{\rho}_1+ ( \rho_2 -  \hat{\rho}_1) m(t)]  \label{hh}.  
\end{eqnarray}
In Eq. (\ref{mm}), $\sigma_2 \approx 0$ and  the terms $(\rho_2 - \hat{\rho}_1)(1 - m(t))$ and $\hat{\sigma}_1(1/h(t) - 1) \ll \alpha_m + \beta_m$ whereas in Eq. (\ref{hh}),
 it may be assumed that $m(t) \approx m_s = \alpha_m/(\alpha_m + \beta_m)$ as m(t) has a faster time constant than h(t).  Therefore, defining
\begin{eqnarray}
\alpha_h    & = & \hat{\sigma}_1 + \sigma_2 \approx \bar{\sigma}_1   \\
\beta_h  & = & \hat{\rho}_1 +  ( \rho_2 -  \hat{\rho}_1) m_s,  
\end{eqnarray}
Eqs. (\ref{mm}) and (\ref{hh}) may be approximated by
\begin{equation}
\frac{dm}{dt}= \alpha_{m}-(\alpha_{m}+\beta_{m})m,  \label{mm1a}
\end{equation}
\begin{eqnarray}
\frac{dh}{dt}= \alpha_{h}-(\alpha_{h}+\beta_{h})h,  \label{hh1a}
\end{eqnarray}
and therefore, during a depolarizing voltage clamp of the Na+ channel membrane, the expression for the open state probability $O(t) = m(t)h(t)$,
where $m(t)$ and $h(t)$ satisfy Eqs.  (\ref{mm1a}) to (\ref{hh1a}), is   in agreement with the voltage clamp solution of Eqs. (\ref{4m1}) to (\ref{4m4})  \cite{vac1}.

Although the K+   and Na+ ion channel currents generate the action potential in the squid axon, it will be assumed that the action potential is determined by the
Na+ and leakage currents, as in myelinated nerve membrane  \cite{crs,bri}, and therefore, the membrane current equation is
\begin{equation}
C\frac{dV}{dt}= I_e -  \bar{g}_{Na} O(t) (V - V_{Na}) - \bar{g}_{L} (V - V_{L}),  \label{cur2}
\end{equation}
\noindent where  the Na+ current is dependent on the open state probability $O(t) = m(t)h(t)$, $\bar{g}_{Na}$ and $\bar{g}_{L}$ are the Na+  and leakage conductances, 
and $V_{Na}$ and $V_{L}$ are the equilibrium potentials, C is the membrane capacitance, and $I_e$ is the electrode current.   The solution of  the master equation for
 coupled activation and inactivation, Eqs.  (\ref{4m1}) to (\ref{4m4}), and Eq. (\ref{cur2}) may be approximated by the solution of Eqs. (\ref{mm1a}), (\ref{hh1a})   and
(\ref{cur2}) (see Fig. 4).
    
  {\bf  Na+ CHANNEL MASTER EQUATION WITH TWO OR THREE ACTIVATION SENSORS}

The time dependence  of the Na+ current during a voltage clamp of myelinated nerve membrane may be described in terms of the expression $m^2 h$ \cite{crs},
and therefore,  assuming that the activation of two voltage sensors  regulating the  Na+ channel  conductance  is coupled to a two-stage inactivation process,
the kinetics is determined by a nine state master equation that may be reduced to a six state system when the first forward and backward inactivation transitions
are rate limiting ($\beta_{ik} \gg \delta_{ik}$ and  $\gamma_{ik} \gg \alpha_{ik}$, for $k=1$ to $3$) \cite{vac1}
\begin{eqnarray}
\frac{dC_1}{dt} & = &  -(\rho_1 + \alpha_C)C_1(t) + \beta_C C_2(t) + \sigma_1 B_1(t) \label{6c1} \\
\frac{dC_2}{dt} & = &  \alpha_C C_1(t) - (\alpha_O + \beta_C + \rho_2) C_2(t) + \beta_O O(t) + \sigma_2 B_2(t) \label{6c2} \\
\frac{dO}{dt}   & = &   \alpha_O C_2(t) - (\beta_O + \rho_3)O(t) +\sigma_3 B_3(t)  \label{6o} \\
\frac{dB_1}{dt} & = &  \rho_1 C_1(t) - (\alpha_{B1} + \sigma_1) B_1(t) + \beta_{B1} B_2(t)  \label{6b1} \\
\frac{dB_2}{dt} & = &  \rho_2 C_2(t) + \alpha_{B1}B_1(t)  - (\alpha_{B2} +\beta_{B1} + \sigma_2) B_2(t) + \beta_{B2} B_3(t)  \label{6b2} \\
\frac{dB_3}{dt} & = &  \rho_3 O(t) + \alpha_{B2}  B_2(t) - (\beta_{B2} + \sigma_3) B_3(t), \label{6b3} 
\end{eqnarray}
 From Eq.  (\ref{6b1}), if $\alpha_{B1} \gg \rho_1 $  and $ \sigma_1 \gg \beta_{B1}$, the occupation probability $B_1(t)$ is dependant on the variation of $C_1(t)$ and $B_2(t)$,
 and therefore, Eqs.  (\ref{6c1}) and (\ref{6b1})  may be reduced to (see Fig. 5)
\begin{eqnarray}
\frac{dC_1}{dt} & = &  -(\alpha_C) + \hat{\rho}_1) C_1(t) + \beta_C C_2(t) + \hat{\sigma}_1 B_2(t) \label{6c1a} \\
\frac{dB_2}{dt} & = &  \hat{\rho}_1  C_1(t)  +  \rho_2 C_2(t)  - (\alpha_{B2} + \hat{\sigma}_1 + \sigma_2) B_2(t) + \beta_{B2} B_3(t)  \label{6b2a} 
\end{eqnarray}
where
\begin{eqnarray}
\hat{\rho}_1    & \approx &  \frac{\rho_1 \alpha_{B1}}{\alpha_{B1} + \sigma_1},  \label{rho1}  \\
\hat{\sigma}_1  & \approx & \frac{\sigma_1 \beta_{B1}  }{\alpha_{B1} + \sigma_1}  \label{sig1}.
\end{eqnarray}

 In Eqs.  (\ref{6b3}) and (\ref{6b2a}),  the $B_2(t)$ and $B_3(t)$ terms are an order of magnitude larger than the closed and open state terms, and therefore, 
defining $B(t) = B_2(t) + B_3(t)$,  the inactivation probabilities $B_2(t)$ and $ B_3(t)$ may be expressed as
\begin{eqnarray}
B_2(t)    & \approx &  \frac{\beta_{B2}  B(t) }{\alpha_{B2} +\beta_{B2}}  \\
B_3(t)   & \approx & \frac{\alpha_{B2}  B(t) }{\alpha_{B2} +\beta_{B2}},
\end{eqnarray}
\noindent and, therefore, Eqs. (\ref{6c1})  to (\ref{6b3})  may be reduced to the four-state master equation (see Fig. 6)
\begin{eqnarray}
\frac{dC_1}{dt} & = &  -(\hat{\rho}_1 + \alpha_C)C_1(t) + \beta_C C_2(t) +\hat{\sigma}_{1r}  B(t)     \label{rr1} \\
\frac{dC_2}{dt} & = &  \alpha_C C_1(t) - (\alpha_O + \beta_C + \rho_2) C_2(t) + \beta_O O(t) + \sigma_{2r}  B(t) \label{rr2} \\
\frac{dO}{dt}   & = &   \alpha_O C_2(t) - (\beta_O + \rho_3)O(t) +  \sigma_{3r} B(t)  \label{rr3} \\
\frac{dB}{dt} & = &  \hat{\rho}_1 C_1(t) + \rho_2 C_2(t) + \rho_3 O(t) - (\hat{\sigma}_{1r} + \sigma_{2r} +  \sigma_{3r} ) B(t)  \label{rr4} 
\end{eqnarray}
where
\begin{eqnarray}
\hat{\sigma}_{1r}   & =  &  \frac{\hat{\sigma}_1  \beta_{B2} }{\alpha_{B2} +\beta_{B2}} \\
\sigma_{2r}          & =  & \frac{\sigma_2 \beta_{B2}  }{\alpha_{B2} +\beta_{B2}}   \\
\sigma_{3r}          & =  &  \frac{\sigma_3   \alpha_{B2} }{\alpha_{B2} +\beta_{B2}},
\end{eqnarray}
\noindent and  $C_1(t) + C_2(t) +O(t) + B_2(t) = 1$.

Assuming that $C_1(t) = m_1(t) h(t)$, $C_2(t) = m_2(t) h(t)$, $O(t) = m_O(t) h(t)$, $B(t) = 1 - h(t)$, where 
$m_1(t), m_2(t)$ and $m_3(t)$ are activation variables and $h(t)$ is an inactivation variable, Eqs. (\ref{rr1}) to (\ref{rr4}), may be expressed as
\begin{eqnarray}
\frac{dm_1}{dt} & = &  -(\hat{\rho}_1 + \alpha_C + \sigma(t) - \rho(t))m_1(t) + \beta_C m_2(t) + \nonumber  \\
                      &    &  \hat{\sigma}_{1r} (1/h(t) - 1)      \label{rm1} \\
\frac{dm_2}{dt} & = &  \alpha_C m_1(t) - (\alpha_O + \beta_C + \rho_2 + \sigma(t) - \rho(t)) m_2(t) + \beta_O m_O(t) + \nonumber  \\
                      &    &  \sigma_{2r}  (1/h(t) - 1) \label{rm2} \\
\frac{dm_O}{dt}   & = &   \alpha_O m_2(t) - (\beta_O + \rho_3 + \sigma(t) - \rho(t))m_O(t) + \nonumber  \\
                      &    &  \sigma_{3r} (1/h(t) - 1)  \label{rm3} \\
\frac{dh}{dt} & = &  (\hat{\sigma}_{1r} + \sigma_{2r} +  \sigma_{3r} )(1 - h(t)) - h(t) \rho(t) 
  \label{rm4} 
\end{eqnarray}
where 
\begin{eqnarray}
\rho(t)  & = &  \hat{\rho}_1 m_1(t) + \rho_2 m_2(t) +  \rho_3 m_O(t)  \label{rhosum} \\
 \sigma(t) & = & ( \hat{\sigma}_{1r}  + \sigma_{2r}  + \sigma_{3r}) (1/h(t) - 1).  \label{sigmasum} 
\end{eqnarray}

As  the inactivation terms are an order of magnitude smaller than the activation terms,   Eqs. (\ref{rm1}) to (\ref{rm3}) may be approximated by
\begin{eqnarray}
\frac{dm_1}{dt} & = &  -\alpha_C m_1(t) + \beta_C m_2(t)      \label{rm1a} \\
\frac{dm_2}{dt} & = &  \alpha_C m_1(t) - (\alpha_O + \beta_C ) m_2(t) + \beta_O m_O(t)  \label{rm2a} \\
\frac{dm_O}{dt}   & = &   \alpha_O m_2(t) - \beta_O m_O(t).  \label{rm3a} 
\end{eqnarray}
The membrane current equation for a Na+ channel with two activation sensors and a leakage channel is
\begin{equation}
C\frac{dV}{dt}= I - \bar{g}_{Na} O(t) (V - V_{Na}) - \bar{g}_{L} (V - V_{L}),  \label{cur3}
\end{equation}
where $O(t) = m_O(t) h(t)$ and the solution of  Eqs. (\ref{6c1}) to (\ref{6b3}) and Eq. (\ref{cur3}) may be approximated by the solution of Eq.  (\ref{rm4}) and  
Eqs. (\ref{rm1a}) to  (\ref{cur3}) (see Fig. 7).

 If the activation sensors are independent ($\alpha_C = 2 \alpha_m = 2 \alpha_O, \beta_O = 2 \beta_C = 2 \beta_m$),  from Eqs.   (\ref{rm1a}) to  (\ref{rm3a}),
$m_1(t) = (1 - m(t))^2 $,  $m_2(t) =  2 m(t)(1 - m(t))$, $m_O(t) = m(t)^2 $,  where $m(t)$ satisfies
\begin{equation}
\frac{dm}{dt}   =  \alpha_m  - m(t)[\alpha_m + \beta_m ]. \label{mm2a} 
\end{equation}
 As the activation variables have a faster time constant than h(t),  $\rho(t)$ may be approximated by  $\beta_h = \hat{\rho}_1 m_{1s} + \rho_2 m_{2s} +  \rho_3 m_{Os}$ where $m_{1s}, m_{2s}$ 
and $m_{Os}$ are the stationary values of  $m_1(t)$,  $m_2(t)$ and  $m_{O}(t)$. Therefore, defining  $\alpha_h = \hat{\sigma}_{1r} + \sigma_{2r} +  \sigma_{3r}$, 
Eq. (\ref{rm4}) may be expressed as 
\begin{equation}
\frac{dh}{dt}  =   \alpha_h - h(t)(\alpha_h + \beta_h), \label{rm4a} 
\end{equation}
and the solution of   Eqs. (\ref{6c1}) to (\ref{6b3})  and Eq.  (\ref{cur3}) may be approximated by the solution of  Eqs.  (\ref{cur3}) to (\ref{rm4a})  (see Fig. 8).
 Assuming that the Na+ channel  conductance is dependent on the activation of three voltage sensors  coupled to a two-stage inactivation process \cite{cgabc},
the kinetics may be described by a twelve state master equation that may be reduced to an eight state system when the first forward and backward inactivation
 transitions are rate limiting  ($\beta_{ik} \gg \delta_{ik}$ and  $\gamma_{ik} \gg \alpha_{ik}$, for $k=1$ to $4$)
\begin{eqnarray}
\frac{dC_1}{dt} & = &  -(\rho_1 + \alpha_{C1})C_1(t) + \beta_{C1} C_2(t) + \sigma_1 B_1(t) \label{8c1} \\
\frac{dC_2}{dt} & = &  \alpha_{C1} C_1(t) - (\alpha_{C2} + \beta_{C1} + \rho_2) C_2(t) + \beta_{C2} C_3(t) + \sigma_2 B_2(t) \label{8c2} \\
\frac{dC_3}{dt} & = &  \alpha_{C2} C_2(t) - (\alpha_{O} + \beta_{C2} + \rho_3) C_3(t) + \beta_{O} O(t) + \sigma_3 B_3(t) \label{8c3} \\
\frac{dO}{dt}   & = &   \alpha_O C_3(t) - (\beta_O + \rho_4) O(t) + \sigma_4 B_4(t)  \label{8o} \\
\frac{dB_1}{dt} & = &  \rho_1 C_1(t) -(\alpha_{B1} + \sigma_1) B_1(t)+\beta_{B1} B_2(t)  \label{8b1} \\
\frac{dB_2}{dt} & = &  \alpha_{B1}B_1(t)  -(\alpha_{B2} +\beta_{B1} + \sigma_2) B_2(t) + \beta_{B2} B_3(t)  + \rho_2 C_2(t) \label{8b2} \\
\frac{dB_3}{dt} & = &    \alpha_{B2}  B_2(t) -(\alpha_{B3} +\beta_{B2} + \sigma_3) B_3(t) + \beta_{B3} B_4(t) + \nonumber  \\
                      &    &    \rho_3 C_3(t),  \label{8b3} \\
\frac{dB_4}{dt} & = &    \alpha_{B3}  B_3(t) - (\beta_{B3} + \sigma_4) B_4(t)  + \rho_4 O(t),  \label{8b4} 
\end{eqnarray}
    From Eq. (\ref{8b1}),  if $\alpha_{B1} \gg \rho_1 $  and $ \sigma_1 \gg \beta_{B1}$, substituting from

\begin{equation}	
B_1(t) \approx   \frac{ \rho_1 C_1(t) +  \beta_{B1} B_2(t)}{\alpha_{B1} + \sigma_1},
\end{equation}
Eqs. (\ref{8c1}) and (\ref{8b1}) may be reduced to
\begin{eqnarray}
\frac{dC_1}{dt} & = &  -(\alpha_{C1} + \hat{\rho}_1) C_1(t) + \beta_{C1} C_2(t) + \hat{\sigma}_1 B_2(t) \label{8c1a} \\
\frac{dB_2}{dt} & = &  \hat{\rho}_1  C_1(t)  +  \rho_2 C_2(t)  - (\alpha_{B2} + \hat{\sigma}_1 + \sigma_2) B_2(t) + \beta_{B2} B_3(t)  \label{8b2a} 
\end{eqnarray}
where $\hat{\rho}_1$ and $\hat{\sigma}_1$ are defined in Eqs. (\ref{rho1}) and (\ref{sig1}). In Eqs. (\ref{8b2})  to  (\ref{8b4}),  the $B_2(t)$,  $ B_3(t)$ and $ B_4(t)$ 
 terms are an order of magnitude larger than the closed and open state terms, and therefore, defining  $B(t) = B_2(t) + B_3(t) + B_4(t)$, the inactivation probabilities
 $B_2(t)$,  $ B_3(t)$ and $ B_4(t)$   may be expressed as
\begin{eqnarray}
B_2(t)    & \approx &  \frac{\beta_{B2} \beta_{B3} B(t) }{\alpha_{B2}\alpha_{B3} + \alpha_{B2}\beta_{B3} +\beta_{B2} \beta_{B3}}   \\
B_3(t)   & \approx & \frac{\alpha_{B2} \beta_{B3} B(t) }{\alpha_{B2}\alpha_{B3} + \alpha_{B2}\beta_{B3} +\beta_{B2} \beta_{B3}} \\
B_4(t)   & \approx & \frac{\alpha_{B2} \alpha_{B3} B(t) }{\alpha_{B2}\alpha_{B3} + \alpha_{B2}\beta_{B3} +\beta_{B2} \beta_{B3}},
\end{eqnarray}
 and Eqs. (\ref{8c1}) to (\ref{8b4}) may be reduced to the five-state master equation (see Fig. 9)  
\begin{eqnarray}
\frac{dC_1}{dt} & = &  -(\rho_1 + \alpha_{C1})C_1(t) + \beta_{C2} C_2(t) +\hat{\sigma}_{1r}  B(t)     \label{ss1} \\
\frac{dC_2}{dt} & = &  \alpha_{C1} C_1(t) - (\alpha_{C2} + \beta_{C1} + \rho_2) C_2(t) + \beta_{C2} C_3(t) + \sigma_{2r}  B(t) \label{ss2} \\
\frac{dC_3}{dt} & = &  \alpha_{C2} C_2(t) - (\alpha_{O} + \beta_{C2} + \rho_3) C_3(t) + \beta_{O} O(t)  + \sigma_{3r}  B(t)  \label{ss3} \\
\frac{dO}{dt}    & = &   \alpha_O C_3(t) - (\beta_O + \rho_4) O(t) +  \sigma_{4r} B(t)  \label{ss4} \\
\frac{dB}{dt}    & = &  \hat{\rho}_1 C_1(t) + \rho_2 C_2(t) + \rho_3 C_3(t) + \rho_4 O(t) -  \nonumber  \\
                      &    &  (\hat{\sigma}_{1r} + \sigma_{2r} +  \sigma_{3r} +  \sigma_{4r}) B(t)  \label{ss5} 
\end{eqnarray}
where 
\begin{eqnarray}
\hat{\sigma}_{1r}   & =  &  \frac{\hat{\sigma}_1  \beta_{B2} \beta_{B3}}{\alpha_{B2}\alpha_{B3} + \alpha_{B2}\beta_{B3} +\beta_{B2}\beta_{B3}} \\
\sigma_{2r}          & =  &  \frac{\sigma_2 \beta_{B2} \beta_{B3}  }{\alpha_{B2}\alpha_{B3} + \alpha_{B2}\beta_{B3} +\beta_{B2}\beta_{B3}}   \\
\sigma_{3r}          & =  &  \frac{\sigma_3   \alpha_{B2} \beta_{B3}}{\alpha_{B2}\alpha_{B3} + \alpha_{B2}\beta_{B3} +\beta_{B2}\beta_{B3}}, \\
\sigma_{4r}          & =  &  \frac{\sigma_4   \alpha_{B2} \alpha_{B3}}{\alpha_{B2}\alpha_{B3} + \alpha_{B2}\beta_{B3} +\beta_{B2}\beta_{B3}},
\end{eqnarray}
\noindent and  $C_1(t) + C_2(t) + C_3(t)+O(t) + B(t) = 1$.

Assuming that $C_1(t) = m_1(t) h(t)$, $C_2(t) = m_2(t) h(t)$, $C_3(t) = m_3(t) h(t)$, $O(t) = m_O(t) h(t)$,$B(t) = 1 - h(t)$, where 
$m_1(t),m_2(t),m_3(t) $ and $m_O(t)$ are activation variables and $h(t)$ is an inactivation variable, Eqs. (\ref{ss1}) to (\ref{ss5}), may be expressed as
\begin{eqnarray}
\frac{dm_1}{dt} & = &  -(\rho_1 + \alpha_{C1} + \sigma(t) - \rho(t))m_1(t) + \beta_{C1} m_2(t) +  \nonumber  \\
                      &    &  \hat{\sigma}_{1r} (1/h(t) - 1)      \label{sm1} \\
\frac{dm_2}{dt} & = &  \alpha_{C1} m_1(t) - (\alpha_{C2} + \beta_{C1} + \rho_2 + \sigma(t) - \rho(t)) m_2(t) + \beta_{C2} m_3(t) + \nonumber  \\
                      &    & \sigma_{2r}  (1/h(t) - 1) \label{sm2} \\
\frac{dm_3}{dt} & = &  \alpha_{C2} m_2(t) - (\alpha_{O} + \beta_{C2} + \rho_3 + \sigma(t) - \rho(t)) m_3(t) + \beta_{O} m_O(t)  + \nonumber  \\
                      &    &  \sigma_{3r}  (1/h(t) - 1) \label{sm3} \\
\frac{dm_O}{dt}   & = &   \alpha_O m_3(t) - (\beta_O + \rho_4 + \sigma(t) - \rho(t)) m_O(t) +  \nonumber  \\
                      &    &  \sigma_{4r} (1/h(t) - 1)  \label{sm4} \\
\frac{dh}{dt} & = & (\hat{\sigma}_{1r} + \sigma_{2r} +  \sigma_{3r} +  \sigma_{4r})(1 - h(t)) - h(t) \rho(t)   \label{sm5} 
\end{eqnarray}
where 
\begin{eqnarray}
\rho(t)  & = &  \hat{\rho}_1 m_1(t) + \rho_2 m_2(t) + \rho_3 m_3(t) +  \rho_4 m_O(t)  \label{rhosum2} \\
 \sigma(t) & = & ( \hat{\sigma}_{1r}  + \sigma_{2r}  + \sigma_{3r} + \sigma_{4r}) (1/h(t) - 1).  \label{sigmasum2} 
\end{eqnarray}
The inactivation terms are an order of magnitude  smaller than the activation terms, and therefore, Eqs. (\ref{sm1}) to (\ref{sm4}) may be approximated by
\begin{eqnarray}
\frac{dm_1}{dt}   & = &  -\alpha_{C1} m_1(t) + \beta_{C1} m_2(t)      \label{sm1a} \\
\frac{dm_2}{dt}   & = & \alpha_{C1} m_1(t) - (\alpha_{C2} + \beta_{C1}) m_2(t) + \beta_{C2} m_3(t)  \label{sm2a} \\
\frac{dm_3}{dt}   & = &  \alpha_{C2} m_2(t) - (\alpha_{O} + \beta_{C2}) m_3(t) + \beta_O m_O(t)  \label{sm3a} \\
\frac{dm_O}{dt}   & = &   \alpha_O m_3(t) - \beta_O m_O(t).  \label{sm4a} 
\end{eqnarray}

  If the activation sensors are independent  ($\alpha_{C1} = 3 \alpha_m, \alpha_{C2} = 2 \alpha_m, \alpha_O =  \alpha_m, \beta_{C1} = 
 \beta_m, \beta_{C2} = 2 \beta_m,\beta_{O} = 3 \beta_m $),      Eqs. (\ref{sm1a}) to (\ref{sm4a}) have the solution 
$m_1(t) = (1 - m(t))^3 $, $m_2(t) =  3 m(t)(1 - m(t))^2$, $m_3(t) =  3 m(t)^2(1 - m(t))$,$m_O(t) = m(t)^3 $,  where $m(t)$ satisfies
\begin{equation}
\frac{dm}{dt}   =  \alpha_m  - m(t)[\alpha_m + \beta_m ]. \label{mm3} 
\end{equation}
 From Eq. (\ref{rhosum2}),  the  inactivation rate $\rho(t)$ in Eq.  (\ref{sm5})  is dependent on the activation variable $m(t)$ as well as the forward inactivation rates. 
However,  as the activation variable $m(t)$ generally has a faster time constant than h(t),   $\rho(t)$ may be approximated by  
\begin{equation}
\beta_h = \hat{\rho}_1 m_{1s} + \rho_2 m_{2s} + \rho_3 m_{3s}+  \rho_4 m_{Os} \label{beth4} 
\end{equation}
where $m_{1s}, m_{2s}, m_{3s}$ and $m_{Os}$ are the stationary values for each membrane potential, and Eq. (\ref{sm5}) may be expressed as 
\begin{equation}
\frac{dh}{dt}        =   \alpha_h - h(t)(\alpha_h + \beta_h)   \label{sm5a} 
\end{equation}
where
\begin{equation}
\alpha_h = \hat{\sigma}_{1r} + \sigma_{2r} +  \sigma_{3r} +  \sigma_{4r}.  \label{alfh4} 
\end{equation}

Assuming that the K+  and leakage channels repolarize the membrane, and the K+ conductance is proportional to $n(t)^4$ where the activation variable    n(t)   satisfies the equation
\begin{equation}
\frac{dn}{dt}   =  \alpha_n  - n(t)[\alpha_n + \beta_n ], \label{nn} 
\end{equation}
and $\alpha_n$ and $\beta_n$ are rate functions,  the membrane current equation is 
\begin{equation}
C\frac{dV}{dt}= I - \bar{g}_{Na} O(t) (V - V_{Na}) - \bar{g}_{K} n(t)^4 (V - V_{K}) - \bar{g}_{L} (V - V_{L}),  \label{cur4}
\end{equation}
where $O(t) = m_O(t) h(t)$, $\bar{g}_{K}$ is the K+  conductance, and $V_{K}$ is the K+ equilibrium potential. The solution of  Eqs. (\ref{8c1}) to (\ref{8b4}), and Eqs. (\ref{nn})  
and Eq. (\ref{cur4}) may be approximated by the solution of Eqs. (\ref{mm3}), (\ref{sm5a}),  (\ref{nn})  and  (\ref{cur4}) - see Figs. 10 and  11 for a Na+ channel with an
 inactivation rate independent of the closed or open state \cite{hh}, and Fig. 12 for a channel where the probability of Na+ inactivation increases with the degree of
activation of the channel \cite{cgabc}. 

   {\bf CONCLUSION}

 Based on an empirical description of the voltage clamp K+ and Na+ channel currents and the calculation of the membrane potential  from the ion current equation, the HH
 model accounts for the shape of the action potential waveform, the speed of propagation, the threshold potential, and the refractory period  of the squid axon membrane \cite{hh}. 
The model assumes that the activation and opening of Na+ channels is independent of the inactivation process that blocks Na+ conductance, and is mathematically equivalent to a
 Markovian master equation with three activation sensors and one inactivation sensor where the inactivation and recovery rate functions are independent of the closed or open state, 
and the activation and deactivation rate functions between closed states are equal to those between inactivated states. However, experimentally, the inactivation rate is dependant on 
the degree of activation \cite{ab}, and the recovery from inactivation is more probable following deactivation \cite{kb}, and thus activation and inactivation are coupled processes. 

   A master equation for coupled activation and two-stage inactivation accounts for the kinetics and voltage dependence of Na+ inactivation and the recovery from inactivation when
the backward inactivation rate is small for the open state but increases as the activation sensors deactivate \cite{cgabc}. From the solution of a nine state master equation with two 
activation sensors during a voltage clamp, the open state probability may be expressed as $m(t)^2 h(t)$  where m(t) and h(t) satisfy rate equations \cite{vac1}, and therefore,
 the HH description of the Na+ current during a voltage clamp is consistent with a coupled Na+ channel gating model.
 
In this paper, it has been shown that a  master equation  that  describes the gating of a Na+ channel with a single activation sensor coupled to inactivation, may be approximated 
by interacting rate equations for inactivation and activation when the first forward and backward transitions are rate limiting. A nine state master equation describing Na+ channel gating with 
two activation sensors and two-stage inactivation may be reduced to a five state system when $\alpha_{B} \gg \rho_1$ and $\sigma_1 \gg \beta_B$ and the first inactivated state
$B_1(t)$ makes a small contribution to the dynamics. For the remaining inactivated state equations,  the $B_2(t)$ and $B_3 (t)$  terms are an order of magnitude larger than
the closed and open state terms, and defining $ B(t) = B_2(t) + B_3 (t)$,  the system may be reduced to kinetic equations in the activation variables $m_1$, $m_2$ and $m_O$ 
and the inactivation variable h. 

If a Na+ channel has three activation sensors and a two-stage inactivation process, a twelve state master equation may be reduced to a system
 of  equations in the variables  $m_1$, $m_2$, $m_3$, $m_O$    and $h$, where the expression for the inactivation rate is dependent on the forward transition rates
 of the DIV sensor as well as the degree of activation of the other sensors, and the rate of recovery from inactivation is dependent on the rate functions of the DIII sensor
 between inactivated states, in agreement with experimental studies and  the known structure of the Na+ channel. By assuming  that the activation sensors are independent,
 the inactivation rate is, in general,  dependent on the activation variable $m(t)$  but,  when  $m(t)$   has a faster time constant than $h(t)$,  it   reduces to a
 voltage-dependent  function,  and therefore,  the  solution of the master equation during an action potential may be approximated by  the  solution of 
HH rate equations for   $m$  and   $h$.

\newpage 

%Figure 1
\begin{figure*}
\begin{center}
\includegraphics[width=0.4 \textwidth]{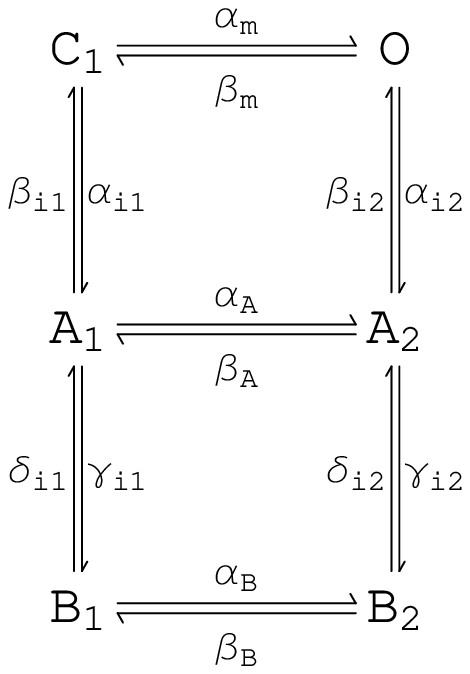}
\caption{
 State diagram for Na+ channel gating  where horizontal transitions represent the activation of a 
single voltage sensor that opens the pore, and vertical transitions represent a two-stage inactivation process.  
}
\end{center}
\end{figure*}

%Figure 2
\begin{figure*}
\begin{center}
\includegraphics[width=0.5\textwidth]{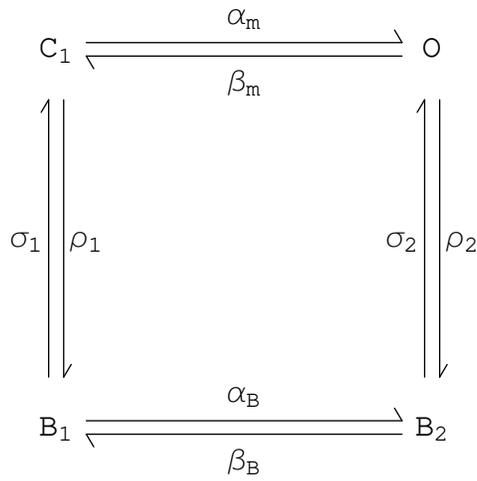}
\caption{
The six state system for Na+ channel gating in Fig. 1 may be reduced to a four state system when 
$\beta_{ik} \gg \delta_{ik}$ and  $\gamma_{ik} \gg \alpha_{ik}$, for $k=1$ to $2$, where $\rho_k$ and 
$\sigma_k$ are derived rate functions for a two-stage Na+ inactivation process. 
}
\end{center}
\end{figure*}

%Figure 3
\begin{figure*}
\begin{center}
\includegraphics[width=0.6 \textwidth]{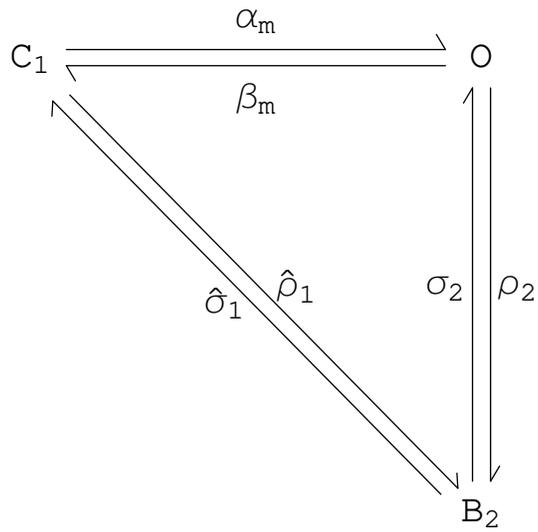}
\caption{ 
The four state system for Na+ channel gating in Fig. 2 may be reduced to a three state system when
$\alpha_{B} \gg \rho_1$  and $ \sigma_1 \gg \beta_{B}$, and $B_1(t) \approx (\rho_1 C_1(t) +  \beta_{B} B_2(t))/(\alpha_{B} + \sigma_1)$.   
}
\end{center}
\end{figure*}

%Figure 4 
\begin{figure*}
\begin{center}
\includegraphics[width=0.5 \textwidth]{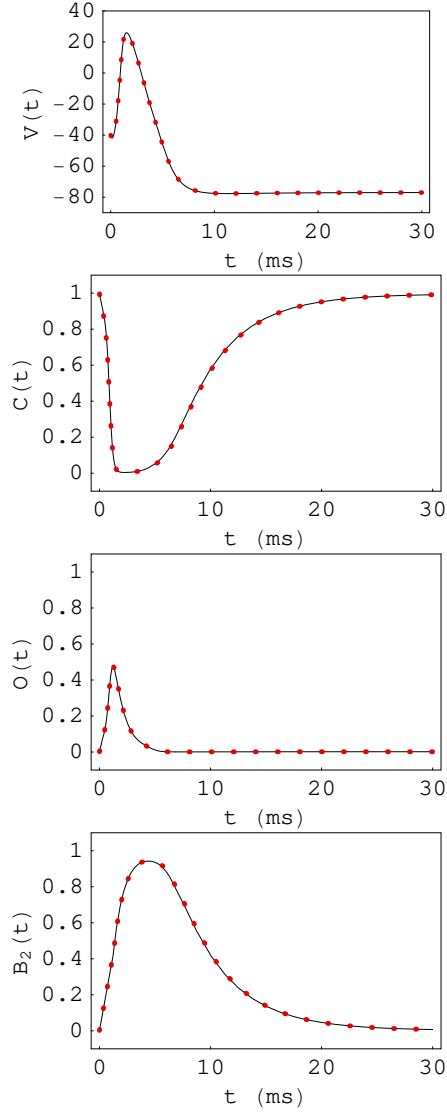}
\caption{
The action potential solution for a master equation describing single sensor activation of a Na+ channel, Eqs.  
(\ref{4m1}) to (\ref{4m4}), and the current equation, Eq. (\ref{cur2}) (solid line) is approximated by the 
solution of rate equations for Na+ activation and inactivation, Eqs. (\ref{mm1a}) and (\ref{hh1a}), and Eq.
(\ref{cur2}) (dotted line), where the rate functions are 
$\alpha_{m} = 0.1(V+25)/(1 - \exp[-(V+25)/10])$, $\beta_m = 4 \exp[-(V+50)/18]$, $ \alpha_{B}=  \alpha_{m}$, 
$\beta_{B} = 0.0135 \beta_{m}$, $ \rho_1 =  \rho_2 = 1/(1+0.17\exp[-2.3V/25])$,  
$\sigma_1 =  2.5/(1+5.9\exp[2.3V/25])$, $\sigma_1 =  0.0135 \sigma_2$ (ms$^{-1}$), $\bar{g}_{Na} = 8$ mS/cm$^2$, 
$\bar{g}_{L} = 0.9$ mS/cm$^2$, $V_{Na}= 55$ mV, $V_{L}= -80$ mV, $I = 1 \mu$A/cm$^2$. 
}
\end{center}
\end{figure*}

%Figure 5 
\begin{figure*}
\begin{center}
\includegraphics[width=0.7\textwidth]{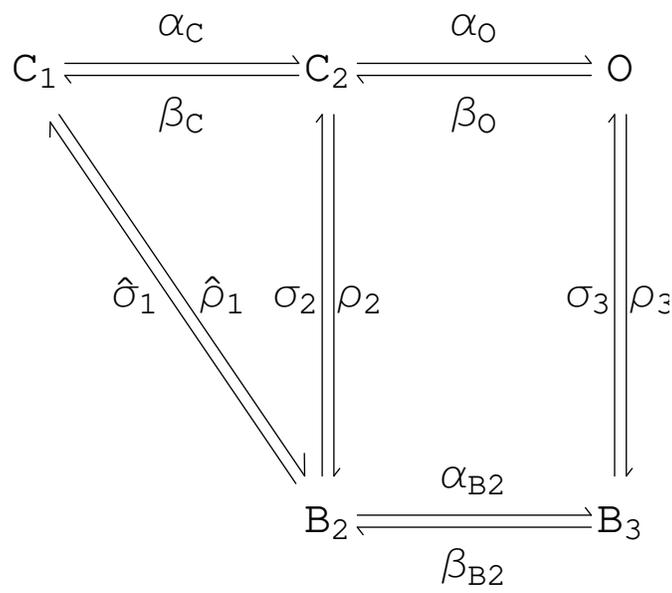}
\caption{
A six-state system for Na+ channel gating may be reduced to a five-state system when $\alpha_{B1} \gg \rho_1$  and  $\sigma_1 \gg \beta_{B1}$.
}
\end{center}
\end{figure*}

%Figure 6 
\begin{figure*}
\begin{center}
\includegraphics[width=0.7\textwidth]{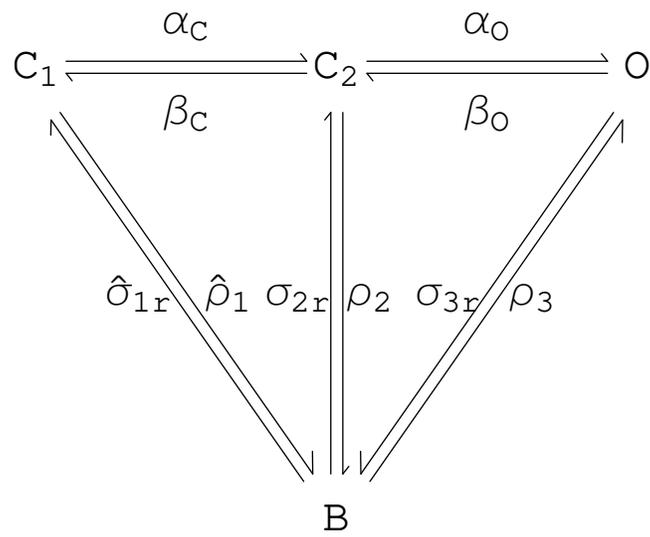}
\caption{
The five state system for Na+ channel gating in Fig. 5 may be reduced to a four state system when the $B_2(t)$ and $B_3(t)$ terms in Eqs.  (\ref{6b3}) and (\ref{6b2a}) are
an order of magnitude larger than the closed and open state terms. 
}
\end{center}
\end{figure*}

%Figure 7 
\begin{figure*}
\begin{center}
\includegraphics[width=0.4 \textwidth]{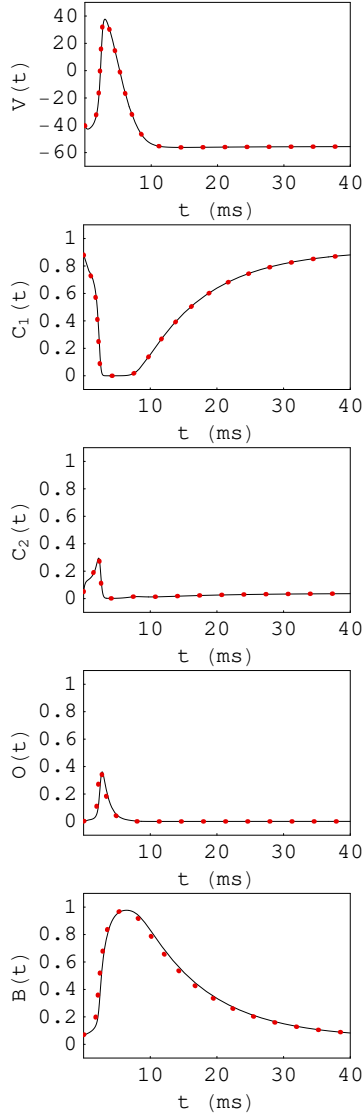}
\caption{
The action potential solution for a Na+ channel six state master equation, Eqs. (\ref{6c1}) to (\ref{6b3})   and the current equation, Eq.  (\ref{cur3}) (solid line) is
approximated by the solution of  Eqs.  (\ref{cur3})  to   (\ref{rm4a}) (dotted line), where the rate functions are
$\alpha_{m} = 0.1(V+25)/(1 - \exp[-(V+25)/10])$, $\beta_m = 4 \exp[-(V+50)/18]$, 
$\alpha_{C} = 2 \alpha_{m }$, $\beta_C = \beta_m$, $\alpha_{O} = \alpha_{m }$, $\beta_O = 2 \beta_m$, 
$\alpha_{B1} =  \alpha_{C}$, $ \beta_{B1} = 0.0135 \beta_C$, $ \alpha_{B2} = \alpha_{O}$, 
$\beta_{B2} = \beta_O$, $\rho_1 =  \rho_2 =  \rho_3 = 1/(1+0.17\exp[-2.3V/25])$,  
$\sigma_1 =  2.5/(1+5.9\exp[2.3V/25])$,$\sigma_2 =  0.0135 \sigma_1$,   $\sigma_3 = \sigma_2$ (ms$^{-1}$),
$g_{Na} = 8$ mS/cm$^2$, $g_{L}=0.9$ mS/cm$^2$, $V_{Na} = 55$ mV, $V_{L}= -80$ mV, $I = 22 \mu$A/cm$^2$.  
}
\end{center}
\end{figure*}

%Figure 8 
\begin{figure*}
\begin{center}
\includegraphics[width=0.4 \textwidth]{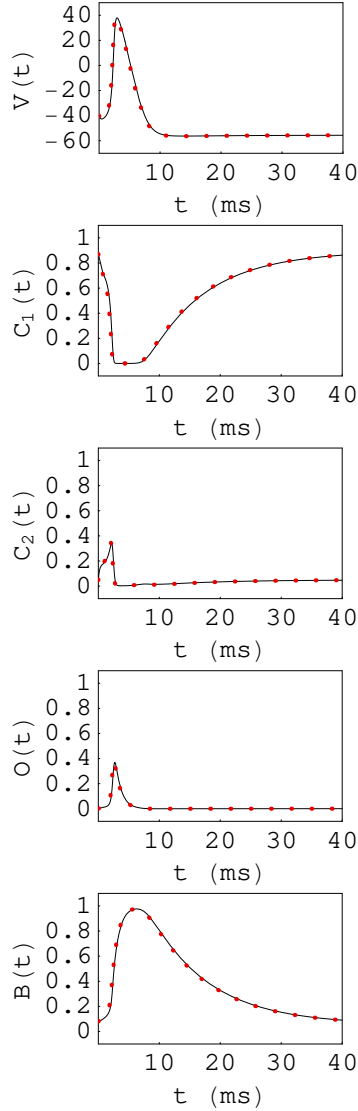}
\caption{
The action potential solution for a Na+ channel six state master equation with two independant activation sensors, Eqs. (\ref{6c1}) to (\ref{6b3})   and the current equation,
Eq.  (\ref{cur3}) (solid line) is approximated by the solution of rate equations for Na+ activation and inactivation and Eq.  (\ref{cur3}) (dotted line), where the rate functions are
 $\alpha_{m} = 0.1(V+25)/(1 - \exp[-(V+25)/10])$, $\beta_m = 4 \exp[-(V+50)/18]$, $\alpha_{C} = 2 \alpha_{m }$, $\beta_C = \beta_m$, $\alpha_{O} = \alpha_{m }$, $\beta_O = 2 \beta_m$, 
$\alpha_{B1} =  \alpha_{C}$, $ \beta_{B1} = 0.0135 \beta_C$, $ \alpha_{B2} = \alpha_{O}$, $\beta_{B2} = \beta_O$, $\rho_1 =  \rho_2 =  \rho_3 = 1/(1+0.17\exp[-2.3V/25])$,  
$\sigma_1 =  2.5/(1+5.9\exp[2.3V/25])$,$\sigma_2 =  0.0135 \sigma_1$,   $\sigma_3 = \sigma_2$ (ms$^{-1}$), $g_{Na} = 8$ mS/cm$^2$, $g_{L}=0.9$ mS/cm$^2$, 
$V_{Na} = 55$ mV, $V_{L}= -80$ mV, $I = 22 \mu$A/cm$^2$.  
}
\end{center}
\end{figure*}

%Figure 9 
\begin{figure*}
\begin{center}
\includegraphics[width=0.8 \textwidth]{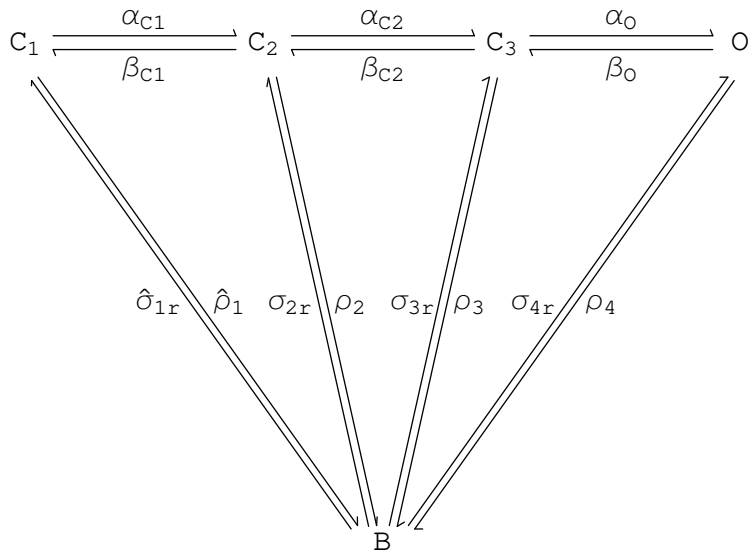}
\caption{
The 8 state system for Na+ channel gating with 3 sensors may be reduced to a 5 state system when $\alpha_{B1} \gg \rho_1$, $ \sigma_1 \gg \beta_{B1}$
and the transition rates between inactivated states are larger than inactivation and recovery rates.
}
\end{center}
\end{figure*}

%Figure 10 
\begin{figure*}
\begin{center}
\includegraphics[width=0.9 \textwidth]{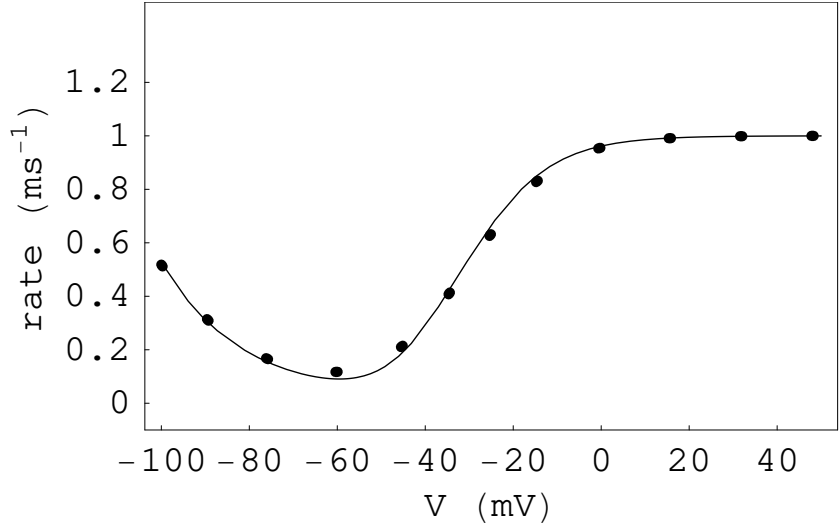}
\caption{
The voltage dependence of the Na+ channel HH inactivation rate function $\alpha_h + \beta_h$, where $\alpha_h = 0.07 \exp[-(V + 60)/20]$  and
$\beta_h = 1/(1+  \exp[-(V + 30)/10])$  may be approximated by the expressions in Eqs.  (\ref{beth4}) and  (\ref{alfh4})   where the rate functions
are defined as $\alpha_{m} = 0.1(V+25)/(1 - \exp[-(V+25)/10])$, $\beta_m = 4 \exp[-(V+50)/18]$,
$\alpha_{C1} = 3 \alpha_{m }$, $\beta_{C1} = \beta_m$, $\alpha_{C2} = 2 \alpha_{m }$, $\beta_{C2} = 2 \beta_m$, $\alpha_{O} = \alpha_{m }$, 
$\beta_O = 3 \beta_m$, $ \alpha_{B1} = 3 \alpha_{C1}$, $ \beta_{B1} = 0.0165 \beta_{C1}$, $ \alpha_{B2} = 2\alpha_{C2}$, $ \beta_{B2} = 2\beta_{C2}$, 
$ \alpha_{B3} = \alpha_{O}$, $ \beta_{B3} = 3\beta_O$,$\rho_1 = \rho_2 = \rho_3 = \rho_4 = 1/(1+0.03 \exp[-2.5V/25])$, $\sigma_1 = 2.5/(1+6\exp[2.5V/25])$, 
$\sigma_2 = 0.0165 \sigma_1$, $\sigma_3 = \sigma_4 = \sigma_2$(ms$^{-1}$), $g_{Na}$= 120 mS/cm$^2$, $g_{K}$= 36 mS/cm$^2$,  $g_{L}$= 0.3 mS/cm$^2$, 
$V_{Na}$ = 55 mV, $V_{L}$ = -50 mV, $I_{e}$ = 9 $\mu$A/cm$^2$.
}
\end{center}
\end{figure*}

%Figure 11
\begin{figure*}
\begin{center}
\includegraphics[width=0.9 \textwidth]{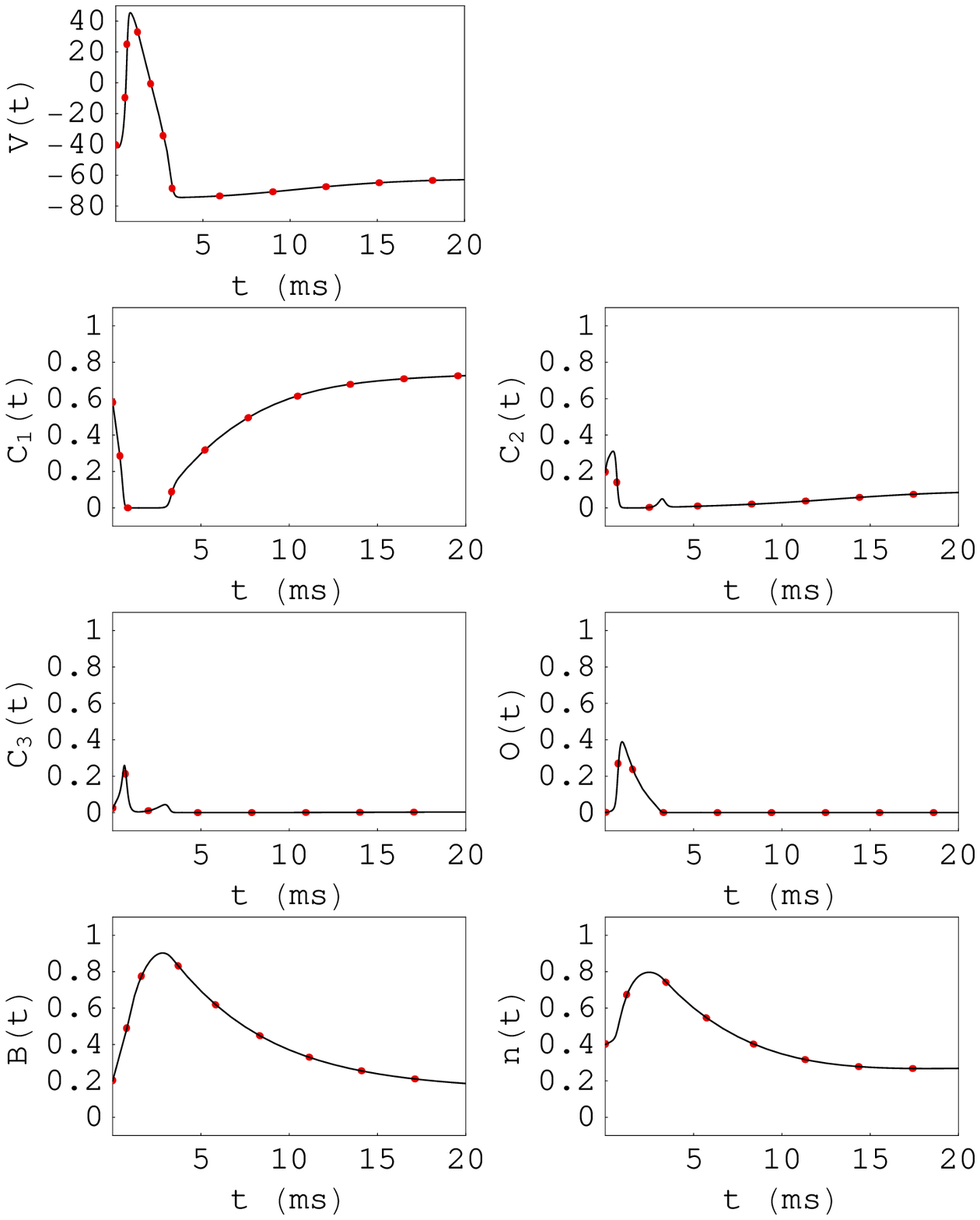}
\caption{ 
The solution of a  Na+ channel eight state master equation, Eqs. (\ref{8c1}) to (\ref{8b4}), Eq. (\ref{nn})  and Eq. (\ref{cur4}) (solid line) may be approximated by the solution
of  Eqs.  (\ref{mm3}) to  Eq.  (\ref{cur4}) (dotted line), where the rate functions are $\alpha_{m} = 0.1(V+25)/(1 - \exp[-(V+25)/10])$,
$\beta_m = 4 \exp[-(V+50)/18]$, $\alpha_{C1} = 3 \alpha_{m }$, $\beta_{C1} = \beta_m$, $\alpha_{C2} = 2 \alpha_{m }$, $\beta_{C2} = 2 \beta_m$, $\alpha_{O} = \alpha_{m }$,
$\beta_O = 3 \beta_m$, $ \alpha_{B1} = 3 \alpha_{C1}$, $ \beta_{B1} = 0.0165 \beta_{C1}$, $ \alpha_{B2} = 2\alpha_{C2}$, $ \beta_{B2} = 2\beta_{C2}$, $ \alpha_{B3} = \alpha_{O}$,
$ \beta_{B3} = 3\beta_O$,$\rho_1 = \rho_2 = \rho_3 = \rho_4 = 1/(1+0.03 \exp[-2.5V/25])$, $\sigma_1 = 2.5/(1+6\exp[2.5V/25])$, 
$\sigma_2 = 0.0165 \sigma_1$, $\sigma_3 = \sigma_4 = \sigma_2$(ms$^{-1}$), $g_{Na}$= 120 mS/cm$^2$, $g_{K}$= 36 mS/cm$^2$, 
$g_{L}$= 0.3 mS/cm$^2$,  $V_{Na}$ = 55 mV, $V_{L}$ = -50 mV, $I_{e}$ = 9 $\mu$A/cm$^2$.
}
\end{center}
\end{figure*}

%Figure 12
\begin{figure*}
\begin{center}
\includegraphics[width=0.9 \textwidth]{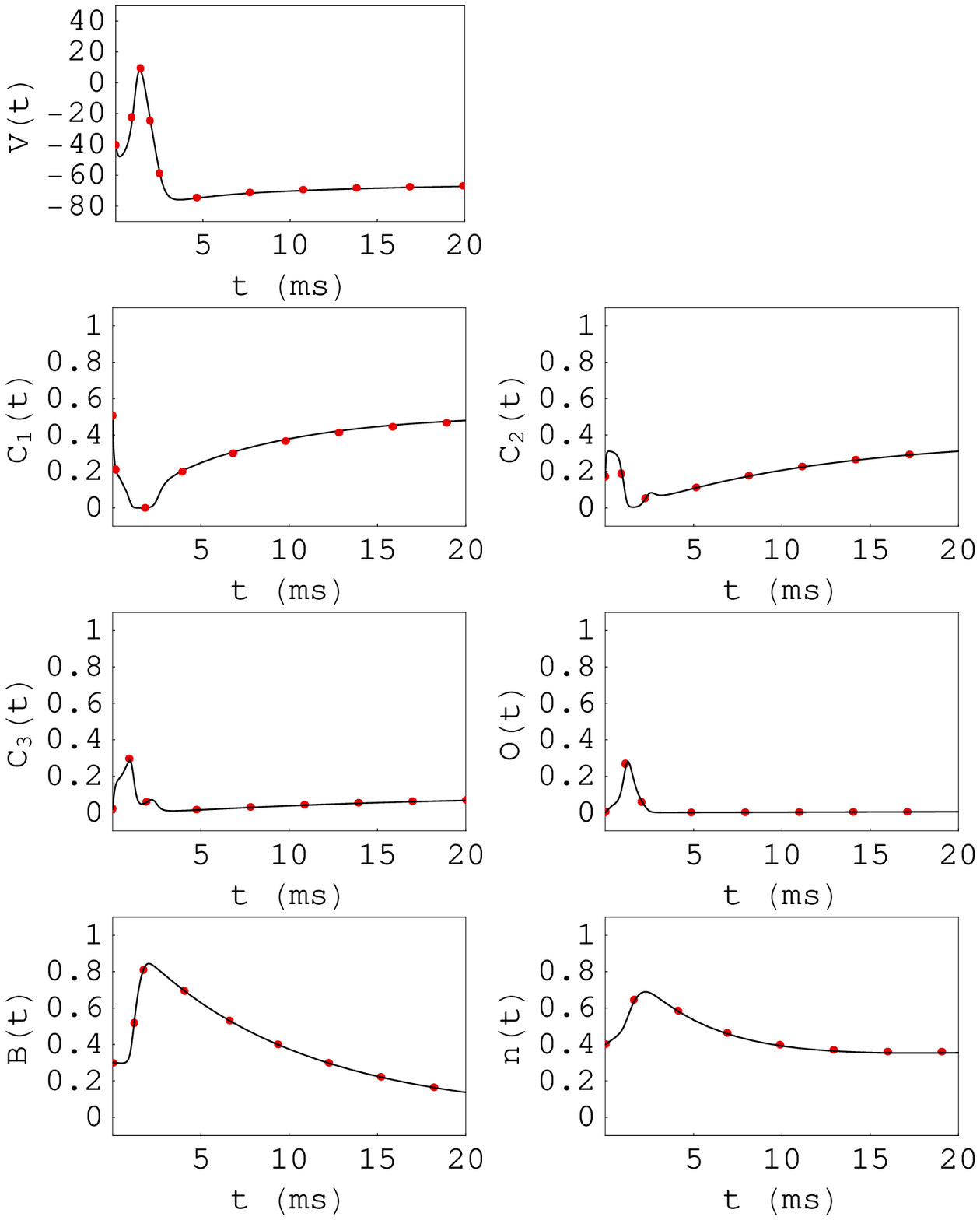}
\caption{ 
The solution of a  Na+ channel eight state master equation, Eqs. (\ref{8c1}) to (\ref{8b4}), Eq. (\ref{nn})  and Eq. (\ref{cur4}) (solid line) may be approximated by the solution
of  Eqs.  (\ref{mm3}) to  Eq.  (\ref{cur4})  (dotted line), where the rate functions are $\alpha_{m} = 7.45 \exp[0.5V/25]$,
$\beta_m = 0.8 \exp[-0.9V/25]$, $\alpha_{C1} = 3 \alpha_{m }$, $\beta_{C1} = \beta_m$, $\alpha_{C2} = 2 \alpha_{m }$, $\beta_{C2} = 2 \beta_m$, $\alpha_{O} = \alpha_{m }$,
$\beta_O = 3 \beta_m$, $ \alpha_{B1} = 3 \alpha_{C1}$, $ \beta_{B1} = 0.01 \beta_{C1}$, $ \alpha_{B2} = 2 \alpha_{C2}$, $ \beta_{B2} = 0.2 \beta_{C2}$, $ \alpha_{B3} = \alpha_{O}$, 
$ \beta_{B3} = 0.3 \beta_O$, $\rho_1 = 2.1/(1+80 \exp[-2.4V/25])$, $\rho_2 = 2.1/(1+8 \exp[-2.4V/25])$, $\rho_3 = 2.1/(1+0.8 \exp[-2.4V/25])$, 
$\rho_4 = 2.1/(1+0.08 \exp[-2.4V/25])$, $\sigma_1 = 80/(80+\exp[2.4V/25])$,  $\sigma_2 = 0.8/(8+\exp[2.4V/25])$, $\sigma_3 = 0.08/(0.8+\exp[2.4V/25])$, 
$\sigma_4 = 0.008/(0.08+\exp[2.4V/25])$ (ms$^{-1}$), $g_{Na}$ = 20 mS/cm$^2$, $g_{K}$ = 6 mS/cm$^2$, $g_{L}$ = 2.3 mS/cm$^2$,  $V_{Na}$ = 55 mV, $V_{L}$  = -80 mV, 
$V_{K}$ = -90 mV, $I_{e}$ = 20 $\mu$A/cm$^2$.
}
\end{center}
\end{figure*}


\newpage 
 
\begin{thebibliography}{99}
\bibitem{hh} A.L. Hodgkin and A.F. Huxley, J. Physiol. \textbf{117}, 500
(1952).

\bibitem{hille} B. Hille, Ion Channels of Excitable Membranes, 3rd ed. (Sinauer, Sunderland, M.A. 2001).

\bibitem{keener} J. Keener, J. Math. Biol. 58, 447 (2009).

\bibitem{ab} C. M. Armstrong, F. Bezanilla, J. Gen. Physiol. \textbf{70}, 567 (1977).

\bibitem{kb} C-C. Kuo and B.P. Bean, Neuron \textbf{12}, 819 (1994).

\bibitem{cgabc}  D.L. Capes, M.P. Goldschen-Ohm, M. Arcisio-Miranda, F. Bezanilla and B. Chanda, 
J. Gen. Physiol. \textbf{142}, 101 (2013).

\bibitem{vac1}  S.R. Vaccaro, Phys. Rev. E \textbf{90}, 052713 (2016).

\bibitem{crs}   S.Y. Chiu, J.M. Ritchie, R.B. Robart and D. Stagg, J. Physiol. \textbf{292}, 149 (1979).

\bibitem{bri}  T. Brismar, J. Physiol. \textbf{298}, 171 (1980).

\end{thebibliography}
\end{document}